\documentclass{llncs}

\ifdefined\pdfminorversion
\fi
\ifdefined\pdfobjcompresslevel
\fi

\usepackage[T1]{fontenc}
\usepackage{lmodern}
\makeatletter
\def\@fnsymbol#1{\ensuremath{\ifcase#1\or *\or\dagger\or\ddagger\else\@ctrerr\fi}}
\makeatother
\usepackage{graphicx}
\usepackage{amsmath,amssymb,mathtools,bm}
\usepackage{booktabs,multirow,tabularx,array,longtable}
\usepackage{xcolor}
\usepackage{float}
\usepackage{listings}
\usepackage{url}
\usepackage{xspace}
\usepackage{seqsplit}
\usepackage{makecell}
\usepackage{pifont}
\usepackage{tikz}
\usetikzlibrary{arrows.meta,positioning,shapes.geometric,fit,calc,backgrounds}
\usepackage{forest}
\usepackage{hyperref}
\usepackage{threeparttable}
\definecolor{themeAcolor1}{HTML}{7B1113}
\definecolor{themeAcolor2}{HTML}{B5192C}
\definecolor{themeAcolor3}{HTML}{D64855}
\definecolor{themeAcolor5}{HTML}{FDF0F1}
\definecolor{themeBcolor1}{HTML}{1A2175}
\definecolor{themeBcolor2}{HTML}{2B37C2}
\definecolor{themeBcolor3}{HTML}{3746F5}
\definecolor{themeBcolor4}{HTML}{C1C5F5}
\definecolor{themeBcolor5}{HTML}{E3E4F5}

\lstdefinestyle{plain}{
  basicstyle=\ttfamily\scriptsize,
  backgroundcolor=\color{white},
  frame=single,
  rulecolor=\color{black},
  framesep=4pt,
  xleftmargin=4pt,
  xrightmargin=4pt,
  breaklines=true,
  numbers=none,
  showstringspaces=false
}
\lstdefinestyle{json}{
  style=plain,
  language={},
  morecomment=[l]{//},
  commentstyle=\itshape\color{gray},
  stringstyle=\color{black}
}

\hypersetup{
  hidelinks,
  pdftitle={SURE-EVAL: A Systematic and Unified Agentic Framework for Reproducible Evaluation}
}

\newcommand{\toolworkflow}{\textsc{Tool Agent Workflow}\xspace}
\newcommand{\mainworkflow}{\textsc{Main Agent Workflow}\xspace}
\newcommand{\detlayer}{\textsc{Deterministic Evaluation Layer}\xspace}
\newcommand{\pass}{\ding{51}}
\newcommand{\fail}{\ding{55}}
\newcommand{\code}[1]{\texttt{\seqsplit{#1}}}
\newcolumntype{P}[1]{>{\raggedright\arraybackslash}p{#1}}
\newcolumntype{Y}{>{\raggedright\arraybackslash}X}
\usepackage{etoolbox}
\makeatletter
\patchcmd{\section}{-18\p@ \@plus -4\p@ \@minus -4\p@}{-8\p@ \@plus -2\p@ \@minus -2\p@}{}{\PackageError{layout}{Section spacing patch failed}{}}
\patchcmd{\section}{12\p@ \@plus 4\p@ \@minus 4\p@}{6\p@ \@plus 2\p@ \@minus 2\p@}{}{\PackageError{layout}{Section spacing patch failed}{}}
\patchcmd{\subsection}{-18\p@ \@plus -4\p@ \@minus -4\p@}{-8\p@ \@plus -2\p@ \@minus -2\p@}{}{\PackageError{layout}{Subsection spacing patch failed}{}}
\patchcmd{\subsection}{8\p@ \@plus 4\p@ \@minus 4\p@}{4\p@ \@plus 2\p@ \@minus 2\p@}{}{\PackageError{layout}{Subsection spacing patch failed}{}}
\patchcmd{\subsubsection}{-18\p@ \@plus -4\p@ \@minus -4\p@}{-10\p@ \@plus -2\p@ \@minus -2\p@}{}{\PackageError{layout}{Run-in spacing patch failed}{}}
\patchcmd{\@maketitle}{\vskip .8cm}{\vskip .4cm}{}{\PackageError{layout}{Title spacing patch failed}{}}
\patchcmd{\abstract}{\advance\topsep by0.35cm}{\advance\topsep by0.1cm}{}{\PackageError{layout}{Abstract spacing patch failed}{}}
\patchcmd{\@maketitle}{\vskip.35cm}{\vskip.2cm}{}{\PackageError{layout}{Author spacing patch failed}{}}
\makeatother
\begin{document}
\pagestyle{empty}
\raggedbottom
\setlength{\abovedisplayskip}{6pt plus 2pt minus 2pt}
\setlength{\belowdisplayskip}{6pt plus 2pt minus 2pt}
\setlength{\abovedisplayshortskip}{3pt plus 1pt}
\setlength{\belowdisplayshortskip}{4pt plus 1pt minus 1pt}

\title{SURE-EVAL: A Systematic and Unified Agentic Framework for Reproducible Evaluation}
\titlerunning{SURE-EVAL}

\author{
Jing Peng\inst{1}\thanks{These authors contributed equally to this work.} \and
Junhao Du\inst{1}\textsuperscript{*}\thanks{Work done during an internship at AISpeech Co., Ltd.} \and
Yixuan Wang\inst{1}\textsuperscript{*}\textsuperscript{\ensuremath{\dagger}} \and
Bowen Wang\inst{3}\textsuperscript{*}\textsuperscript{\ensuremath{\dagger}} \and
Hanqi Li\inst{1}\textsuperscript{*} \and
Chaolei Liu\inst{1} \and
Weihan Chen\inst{1} \and
Haohui Xie\inst{1} \and
Ruichen Sun\inst{6}\textsuperscript{\ensuremath{\dagger}} \and
Chenghao Wang\inst{7}\textsuperscript{\ensuremath{\dagger}} \and
Wen Wen\inst{1} \and
Guanyu Chen\inst{1} \and
Xiaoyu Gu\inst{1} \and
Haoyu Li\inst{1} \and
Yiwei Guo\inst{1} \and
Bohan Li\inst{1} \and
Tao Liu\inst{1} \and
Yucheng Wang\inst{4} \and
Yu Xi\inst{1} \and
Yihua Zhou\inst{2} \and
Qiang Zhou\inst{2} \and
Feng Lu\inst{2} \and
Shuai Wang\inst{5} \and
Kai Yu\inst{1}\textsuperscript{\ensuremath{\ddagger}}}
\authorrunning{J. Peng et al.}
\institute{
\textsuperscript{1}Shanghai Jiao Tong University \quad \textsuperscript{2}AISpeech Co., Ltd.\\
\textsuperscript{3}University of Bristol \quad \textsuperscript{4}ETH Zurich \quad \textsuperscript{5}Nanjing University\\
\textsuperscript{6}Imperial College London \quad \textsuperscript{7}Xi'an Jiaotong University\\[3pt]
\textsuperscript{\ensuremath{\ddagger}}Corresponding author: Kai Yu, \email{kai.yu@sjtu.edu.cn}}

\maketitle
\thispagestyle{empty}

\begin{abstract}
Audio and speech models are being released at an accelerating pace, yet their reported scores often conflate model capability with hidden choices in deployment and evaluation. The same checkpoint can produce different predictions under different runtimes, hardware, numerical precision, decoding parameters, fallback policies, or retry strategies; even fixed predictions can receive different scores when output canonicalization, text normalization, and metric implementations differ. Existing speech benchmarks standardize selected datasets or scoring procedures, but generally do not connect heterogeneous model onboarding, controlled inference, and versioned scoring within one executable workflow. We introduce \textbf{SURE-EVAL}, a \textbf{S}ystematic and \textbf{U}nified Agentic framework for \textbf{R}eproducible \textbf{E}valuation of audio and speech systems. A \toolworkflow converts heterogeneous model releases into isolated, verified callable tools. A \mainworkflow then commits task-specific inference and scoring protocols before delegating every score-bearing operation to versioned deterministic programs. Each result is accompanied by the runtime, protocol, pipeline nodes, predictions, and audit artifacts that produced it. Across 18 public releases spanning Automatic Speech Recognition (ASR), Text-to-Speech (TTS), Voice Conversion (VC), Speaker Diarization (SD), Speaker-attributed ASR (SA-ASR), and multi-task audio understanding, a Codex-only baseline completes 12 models in one shot, whereas the same Codex agent equipped with the SURE-EVAL Tool Agent Workflow completes all 18 in one shot. We further conduct unified evaluations over seven ASR test conditions and two TTS subsets. A compact protocol analysis on three TTS systems finds absolute differences of 0.02--0.52 points between paper-reported and unified results, with the direction varying by model and language. These results show that reproducible evaluation requires controlling both how a model is executed and how its outputs are scored.
\end{abstract}

\section{Introduction}
\label{sec:introduction}

Rapid advances in speech and audio foundation models have accelerated the release of systems for automatic speech recognition (ASR), speech understanding, speaker analysis, text-to-speech synthesis (TTS), voice conversion (VC), and audio-input instruction following. In this paper, audio understanding refers to speech-centered audio-input models that may combine recognition, reasoning, instruction following, and generation within one release. New checkpoints and repositories now appear faster than researchers and practitioners can independently integrate and evaluate them. However, the increased availability of models has not been matched by an equally scalable evaluation infrastructure. In practice, results reproduced from public releases can deviate substantially from the numbers reported in papers or model cards, and scores attached to nominally identical models are often not directly comparable.

The first source of this discrepancy is \emph{execution variance}. A checkpoint does not uniquely identify the system that is evaluated. Its behavior also depends on the runtime and hardware, numerical precision and quantization, checkpoint revision, model wrapper, language and task modes, decoding temperature, beam-search configuration, fallback schedule, retry policy, prompt or context propagation, and streaming or chunking strategy. A reported score may be valid for its original configuration, yet it cannot be assumed to transfer to another implementation when these conditions are incomplete or implicit.

\begin{figure}[!htbp]
\centering
\includegraphics[width=0.82\linewidth]{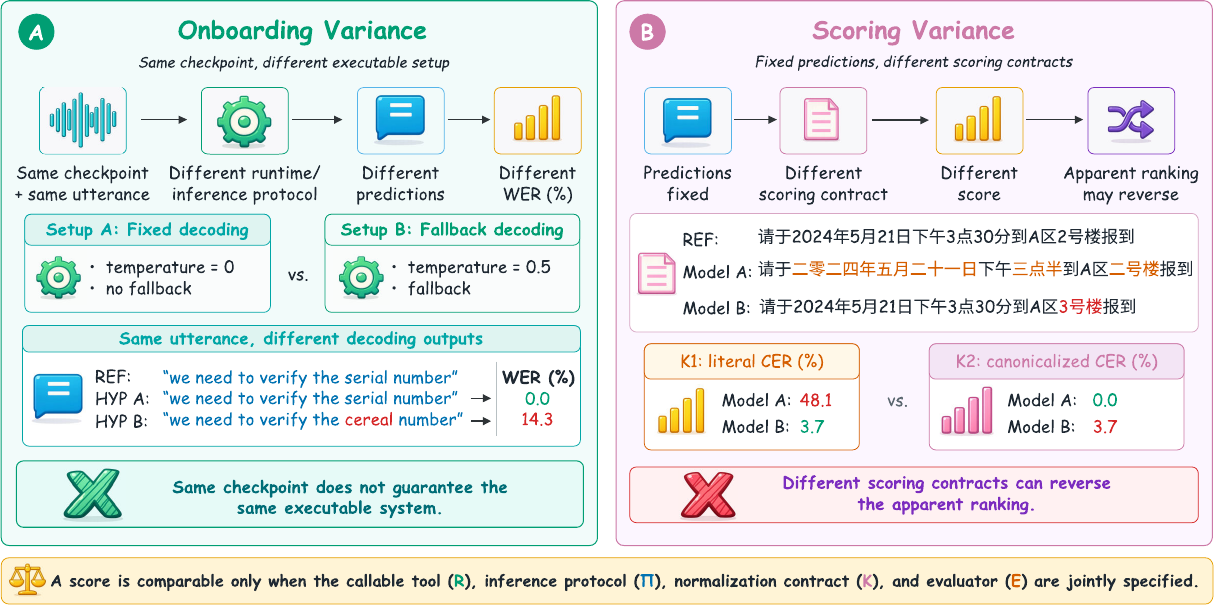}
\caption{Motivation for SURE-EVAL. Comparable scores require jointly specified model onboarding, inference protocol, output normalization, and metric evaluation.}
\label{fig:motivation}
\end{figure}

The second source is \emph{scoring variance}. After inference, a raw model output must be extracted, validated, canonicalized, normalized, tokenized, scored, and aggregated. Omitting a normalization stage can change ASR WER or CER; TTS intelligibility depends on the selected ASR frontend; and two procedures both described as ``normalization'' may implement different equivalence rules or different versions of the same rules. Consequently, fixed predictions can receive different scores, and apparent model rankings may reflect evaluation-pipeline choices rather than model capability. Figure~\ref{fig:motivation} summarizes these two coupled sources of variance.

Speech benchmarks have made important progress toward standardization. SUPERB fixes a broad task suite for comparing reusable speech representations~\cite{yang2021superb}; the Open ASR Leaderboard standardizes multilingual and long-form ASR datasets, normalization, WER, and efficiency reporting~\cite{srivastav2025openasr}; and SURE provides unified data, prediction, normalization, and scoring interfaces for speech understanding~\cite{peng2026sure}. These efforts establish trustworthy benchmark surfaces, but they generally assume that a model has already been made callable under an appropriate inference configuration. Extending them to newly released repositories, incompatible environments, generation tasks, and model-specific decoding controls still requires substantial engineering and often leaves the effective execution protocol under-specified.

We introduce \textbf{SURE-EVAL}, a \textbf{S}ystematic and \textbf{U}nified Agentic framework for \textbf{R}eproducible \textbf{E}valuation. Its central premise is that a score is not an intrinsic property of a checkpoint; it is the outcome of an \emph{executable evaluation claim} that binds a verified callable model, an inference protocol, a canonical dataset, an output contract, a versioned canonicalization route, and a metric evaluator. SURE-EVAL uses agents only where heterogeneous evidence must be interpreted. The \toolworkflow reads repositories, checkpoints, documentation, and execution failures to construct a verified callable tool. The \mainworkflow converts an evaluation request into a committed run whose inference and scoring choices are explicit. Deterministic scripts then perform all score-bearing operations. Thus, agents interpret incomplete evidence, while executable protocols govern the scientific result.

We evaluate this design rather than presenting it only as an engineering specification. The Tool Agent is applied to 18 heterogeneous speech releases spanning recognition, generation, conversion, speaker diarization, speaker-attributed ASR, and multi-task audio systems. The latter includes audio-input foundation models such as Kimi-Audio, which expose broader audio understanding or instruction-following interfaces while still requiring speech-oriented execution and scoring contracts. We compare the same Codex agent with and without the structured Tool Agent workflow, isolating the contribution of context routing, memory discipline, explicit contracts, and staged validation to first-pass completion. The Main Agent is then used to evaluate representative ASR and TTS models under controlled hardware, inference, dataset, and scoring conditions. Finally, a compact protocol analysis compares paper-reported results with values obtained under the unified contract, exposing discrepancies that cannot be interpreted without the corresponding execution and scoring records.

Our contributions are threefold:
\begin{enumerate}
    \item We formulate speech evaluation as an executable claim governed by two coupled contracts: an execution contract that fixes model onboarding and inference conditions, and a scoring contract that fixes output canonicalization and metric computation.
    \item We introduce SURE-EVAL, which combines evidence-guided agentic onboarding, protocol-committed evaluation planning, and versioned deterministic metric pipelines across speech recognition, generation, conversion, and speaker-aware tasks.
    \item We validate the framework on 18 heterogeneous releases and controlled ASR/TTS evaluations, reporting a Codex-only versus Tool-Agent first-pass comparison, unified benchmark results, and a protocol-level analysis of paper-reported versus reproduced scores.
\end{enumerate}

\section{Related Work}
\label{sec:related-work}

\subsection{Reproducible Speech Benchmarking}

SUPERB standardizes lightweight downstream evaluation over frozen speech representations and a broad task suite~\cite{yang2021superb}. Its contribution is a consistent representation-learning benchmark rather than deployment of arbitrary end-to-end model releases. The Open ASR Leaderboard evaluates more than 60 systems over multilingual, short-form, and long-form ASR, with common normalization, WER, and efficiency reporting~\cite{srivastav2025openasr}. It substantially improves ASR transparency, but remains an ASR-specific benchmark with a predefined evaluation surface.

SURE targets speech understanding and organizes datasets, prediction formats, normalization, and scoring across multiple dimensions~\cite{peng2026sure}. SURE-EVAL addresses a complementary layer. It starts from heterogeneous model releases, extends the scope to generation and conversion, and treats runtime, hardware, decoding constraints, output canonicalization, and evaluator versions as one auditable contract. Moreover, adding a new release to SURE-EVAL is an agentic onboarding problem followed by deterministic evaluation, rather than a manually integrated benchmark entry.

\subsection{Agentic Deployment and Evaluation Automation}

Repo2Run uses an LLM agent to infer dependencies and generate executable Docker environments for Python repositories~\cite{repo2run2025}. It targets repository executability, but does not define task-specific inference conditions or score semantics. One-Eval converts natural-language requests into traceable LLM benchmark workflows, including dataset resolution, schema mapping, metric selection, and reporting~\cite{shen2026oneeval}. Its focus is LLM benchmark orchestration after an evaluation interface is available. More general systems such as HELM emphasize explicit scenarios and controlled measurement interfaces~\cite{liang2022helm}, while reproducibility programs emphasize transparent artifacts and reporting~\cite{pineau2020improving}.

SURE-EVAL bridges the boundary between these lines of work: it first converts a heterogeneous speech release into a verified callable system and then evaluates it under an explicit, versioned inference-and-scoring contract. Table~\ref{tab:related-work} summarizes the distinction.

\begin{table}[!htbp]
\setlength{\belowcaptionskip}{3pt}
\centering
\scriptsize
\caption{Scope of representative benchmark and automation frameworks. ``Bounded'' denotes a standardized interface within a predefined benchmark surface.}
\label{tab:related-work}
\begin{tabularx}{\linewidth}{@{}P{22mm}P{27mm}P{25mm}Y@{}}
\toprule
\textbf{Framework} & \textbf{Primary scope} & \textbf{Model onboarding} & \textbf{Inference and scoring control} \\
\midrule
SUPERB~\cite{yang2021superb} & Speech representation benchmark & Bounded model interface & Fixed task recipes \\
Open ASR~\cite{srivastav2025openasr} & Multilingual/long-form ASR & Bounded ASR interface & Unified ASR normalization, WER, and efficiency \\
SURE~\cite{peng2026sure} & Speech understanding & Assisted/manual integration & Unified prediction, normalization, and scoring \\
Repo2Run~\cite{repo2run2025} & Generic Python repositories & Agentic environment construction & No task-specific score contract \\
One-Eval~\cite{shen2026oneeval} & LLM benchmark orchestration & Benchmark resolution & Traceable benchmark-specific metrics \\
\textbf{SURE-EVAL} & Recognition, generation, conversion, and speaker tasks & Agentic release-to-tool construction & Explicit inference contract and versioned scoring nodes \\
\bottomrule
\end{tabularx}
\end{table}

\section{SURE-EVAL Framework}
\label{sec:framework}

\begin{figure}[!htbp]
\centering
\includegraphics[width=0.90\linewidth]{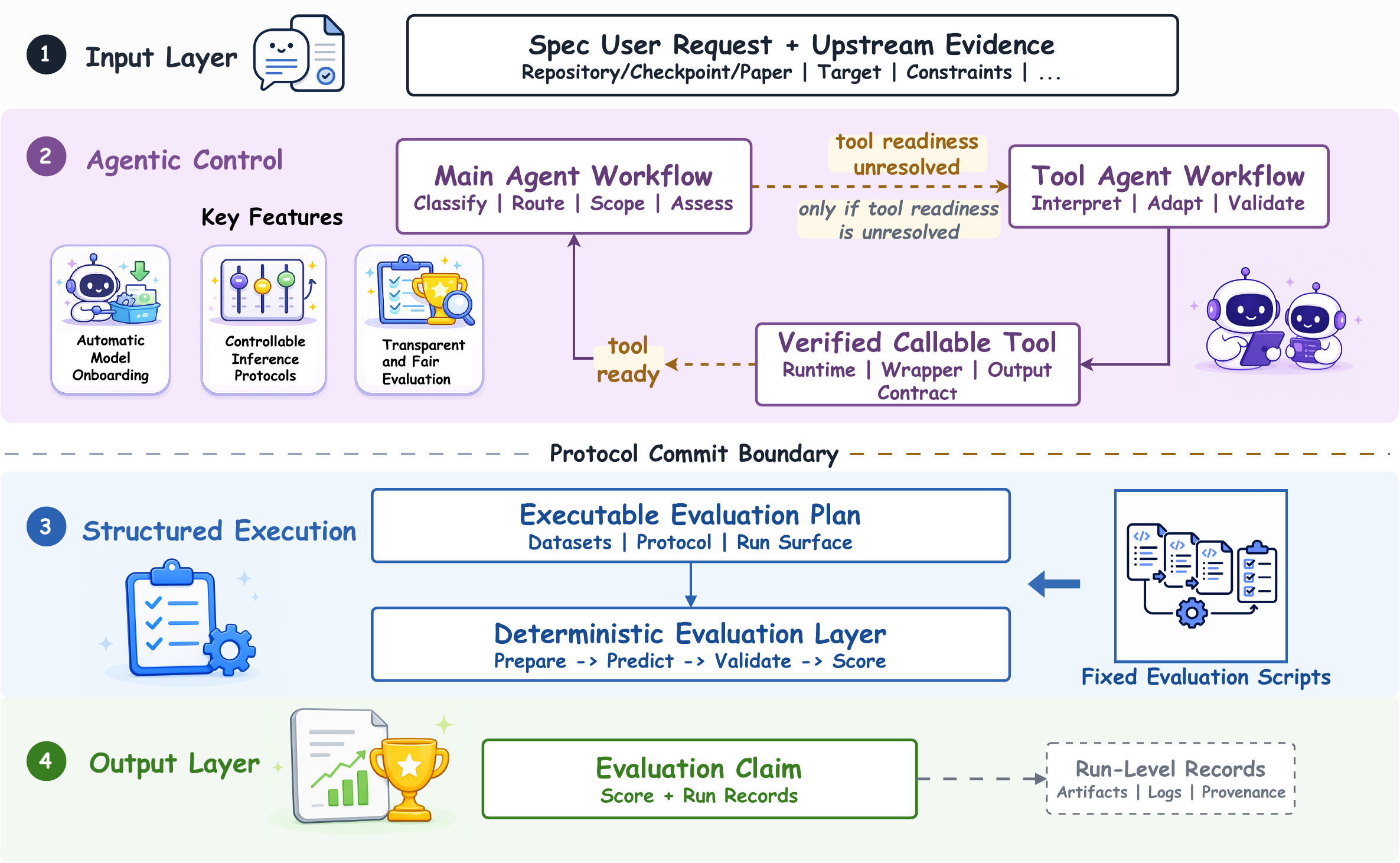}
\caption{Overall SURE-EVAL architecture. The \mainworkflow routes requests, invokes the \toolworkflow when callable readiness is unresolved, and hands score-bearing work to the \detlayer.}
\label{fig:architecture}
\end{figure}

SURE-EVAL connects a user request and heterogeneous upstream evidence to an inspectable result through three components: the \mainworkflow, the \toolworkflow, and the \detlayer. As illustrated in Fig.~\ref{fig:architecture}, the Main Agent is the global entry point. It checks whether a verified callable tool exists and conditionally invokes the Tool Agent when readiness is unresolved. Once the tool is ready, the Main Agent commits the evaluation scope and protocol. All operations that can change the reported score are then executed by registered deterministic programs.

\subsection{Executable Evaluation Claims}
\label{sec:evaluation-contract}

The two sources of variance in Fig.~\ref{fig:motivation} are represented by separate but coupled contracts. The execution contract is
\begin{equation}
\mathcal{C}_{\mathrm{exec}}
=
\left\langle M,R,H,Q,\Pi_{\mathrm{infer}}\right\rangle,
\label{eq:exec-contract}
\end{equation}
where $M$ is the model and checkpoint revision, $R$ is the verified runtime and wrapper, $H$ is the hardware/topology, $Q$ records precision and quantization, and $\Pi_{\mathrm{infer}}$ fixes task-specific inference choices. The scoring contract is
\begin{equation}
\mathcal{C}_{\mathrm{score}}
=
\left\langle D,O,\mathcal{K},\mathcal{E}\right\rangle,
\label{eq:score-contract}
\end{equation}
where $D$ is the canonical dataset instance, $O$ is the output schema and extraction rule, $\mathcal{K}$ is the canonicalization/normalization route, and $\mathcal{E}$ is the metric implementation and aggregation procedure.

A reported result is therefore represented as
\begin{equation}
\begin{aligned}
\Gamma
&=
\left\langle
\mathcal{C}_{\mathrm{exec}},
\mathcal{C}_{\mathrm{score}},
 s,
\mathcal{A}
\right\rangle,\\
s
&=\mathcal{E}\!\left(
\mathcal{K}\!\left(
O\!\left(\operatorname{Predict}(M,R,H,Q,\Pi_{\mathrm{infer}},D)\right)
\right)
\right).
\end{aligned}
\label{eq:evaluation-claim}
\end{equation}
where $s$ is the score and $\mathcal{A}$ is the run-level audit record. This formulation prevents a scalar from being detached from the executable conditions that produced it.

The architectural control principle follows directly: agents may interpret natural-language requests, repository documentation, error traces, and incomplete evidence, but a decision that changes score semantics must be materialized as a structured protocol input before execution. Agents select among declared alternatives; they do not write normalization logic, substitute evaluators, or improvise score-bearing commands during a run. This separation follows the protocol-programming view that semantic reasoning may remain inside actors while enforceable commitments belong in the execution harness~\cite{xflow2026}.

The contract is also the unit of comparison and reuse. Two results are directly comparable only when the relevant execution and scoring fields are equal or belong to a declared compatible protocol class. Conversely, a changed checkpoint, precision mode, decoding policy, normalizer version, or evaluator revision yields a distinct claim even when the displayed model family and metric name are unchanged. SURE-EVAL therefore preserves both a human-readable protocol summary and machine-readable resolved identifiers. This makes protocol differences inspectable before execution and prevents later registry updates from retroactively changing the interpretation of an archived score.

\subsection{Tool Agent: Evidence-Guided Model Onboarding}
\label{sec:toolworkflow}

The \toolworkflow establishes a verified callable tool or returns a documented non-adaptation verdict. It begins from a bounded onboarding hypothesis containing the model identity, repository/checkpoint pointers, runtime envelope, and callable input/output contract. Unknown fields remain explicit: an uncertain weight location or device requirement must be resolved from evidence, preserved as unresolved, or surfaced as a blocking condition. The Tool Agent does not select evaluation datasets, tune benchmark settings, or compute scores.

\begin{figure}[!htbp]
\centering
\includegraphics[width=0.82\linewidth]{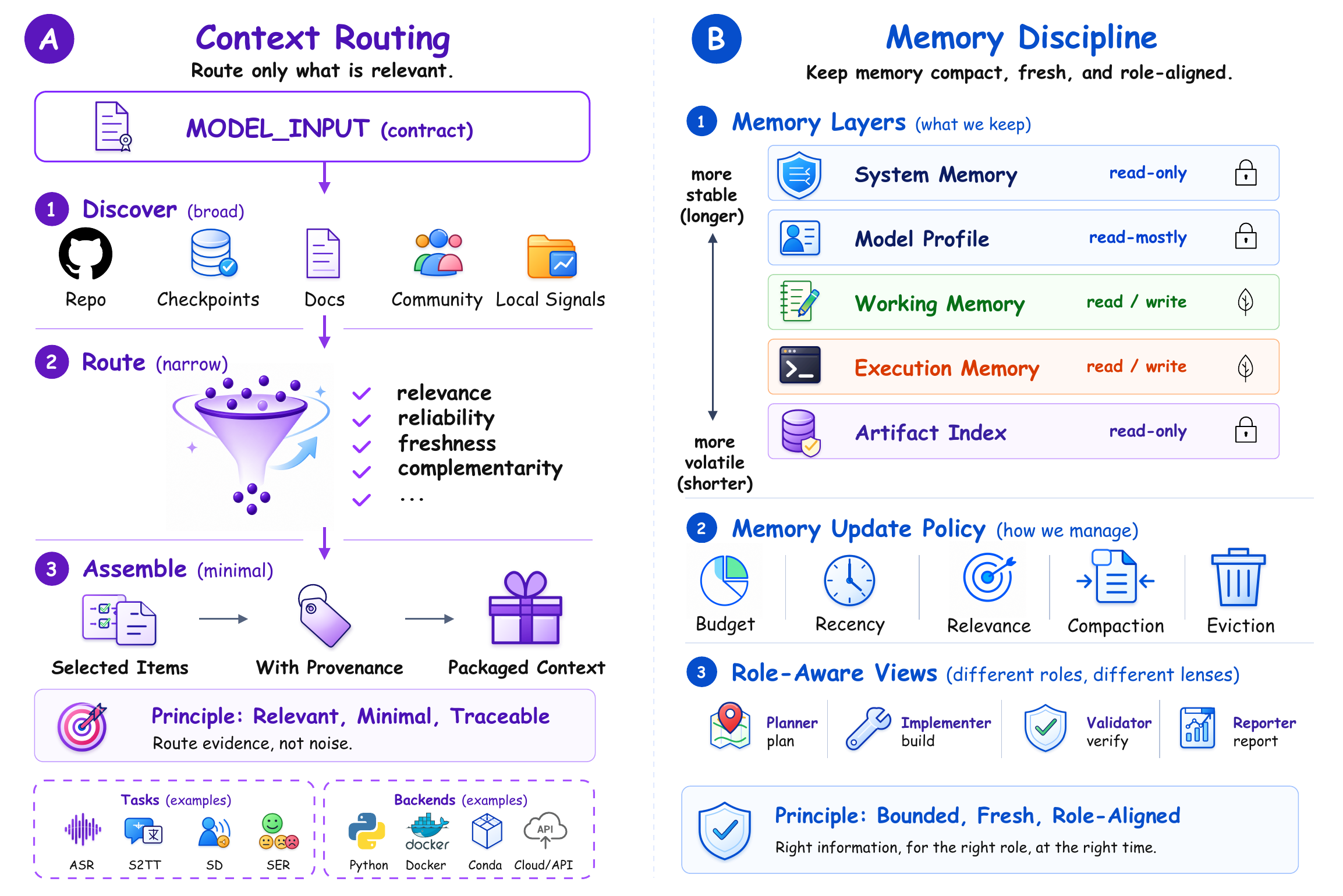}
\caption{Tool Agent Workflow module view. Context routing selects task, environment, and failure-specific knowledge while preserving bounded memory and auditable evidence.}
\label{fig:tool-agent-module}
\end{figure}

The distinctive mechanism is selective context routing with bounded memory, shown in Fig.~\ref{fig:tool-agent-module}. Discovery first gathers repository files, checkpoints, documentation, community evidence, and local execution signals. Routing then selects a minimal context according to relevance, reliability, freshness, and complementarity. Task-specific guidance covers ASR, speech translation, speaker diarization, speech understanding, TTS, VC, and related interfaces; environment-specific guidance covers Python package managers, containers, and APIs. Failure-specific records are loaded only after an observed error or ambiguity, preventing unrelated historical failures from polluting a straightforward installation.

Memory is separated into system rules, a read-mostly model profile, volatile working and execution memories, and an artifact index. Role-aware views expose different slices to the planner, implementer, validator, and reporter. Compaction and eviction keep the active context bounded while preserving provenance. This design is important for model onboarding because repository evidence is heterogeneous and rapidly changing, whereas the callable contract must remain concise and stable.

This routing discipline turns repository adaptation into an evidence-selection problem rather than an unrestricted coding session. The planner receives source identity, deployment constraints, and unresolved fields; the implementer receives only the selected build and wrapper evidence; the validator receives the callable contract and produced artifacts; and the reporter receives validation outcomes and provenance. Separating these views reduces two common failure modes of generic coding agents: repeatedly revisiting already rejected installation paths and allowing a local workaround to weaken the required interface. Any source edit, environment choice, or fallback path is therefore attached to the model-local artifact record rather than retained only in conversational context.

For local releases, the selected evidence is converted into an isolated model-local environment and wrapper; remote APIs are encapsulated behind the same callable interface when permitted. A candidate tool must pass four ordered gates: \emph{import}, \emph{model/weight loading}, \emph{minimal inference}, and \emph{output-contract validation}. A successful process with a malformed output is not evaluation-ready. Only after all gates pass does the workflow return the model specification, runtime entry point, validation verdict, sample output, and artifact manifest to the Main Agent. Failures are classified and repaired inside onboarding rather than being silently absorbed by the later benchmark.

The gates form a monotonic readiness boundary. Import success establishes only that the package graph is loadable; load success additionally resolves the intended checkpoint and device path; inference success demonstrates that the selected entry point is callable on a bounded fixture; and contract success verifies that the returned fields, files, and metadata can enter deterministic evaluation without model-specific interpretation. A later gate cannot compensate for an earlier failure, and the workflow is not allowed to relax the requested output schema merely to declare success. The handoff is consequently a verified executable interface, not a repository directory or an informal statement that the model ``runs.

\subsection{Main Agent: Committing Comparable Evaluation}
\label{sec:main-agent}

\begin{figure}[!htbp]
\centering
\includegraphics[width=0.90\linewidth]{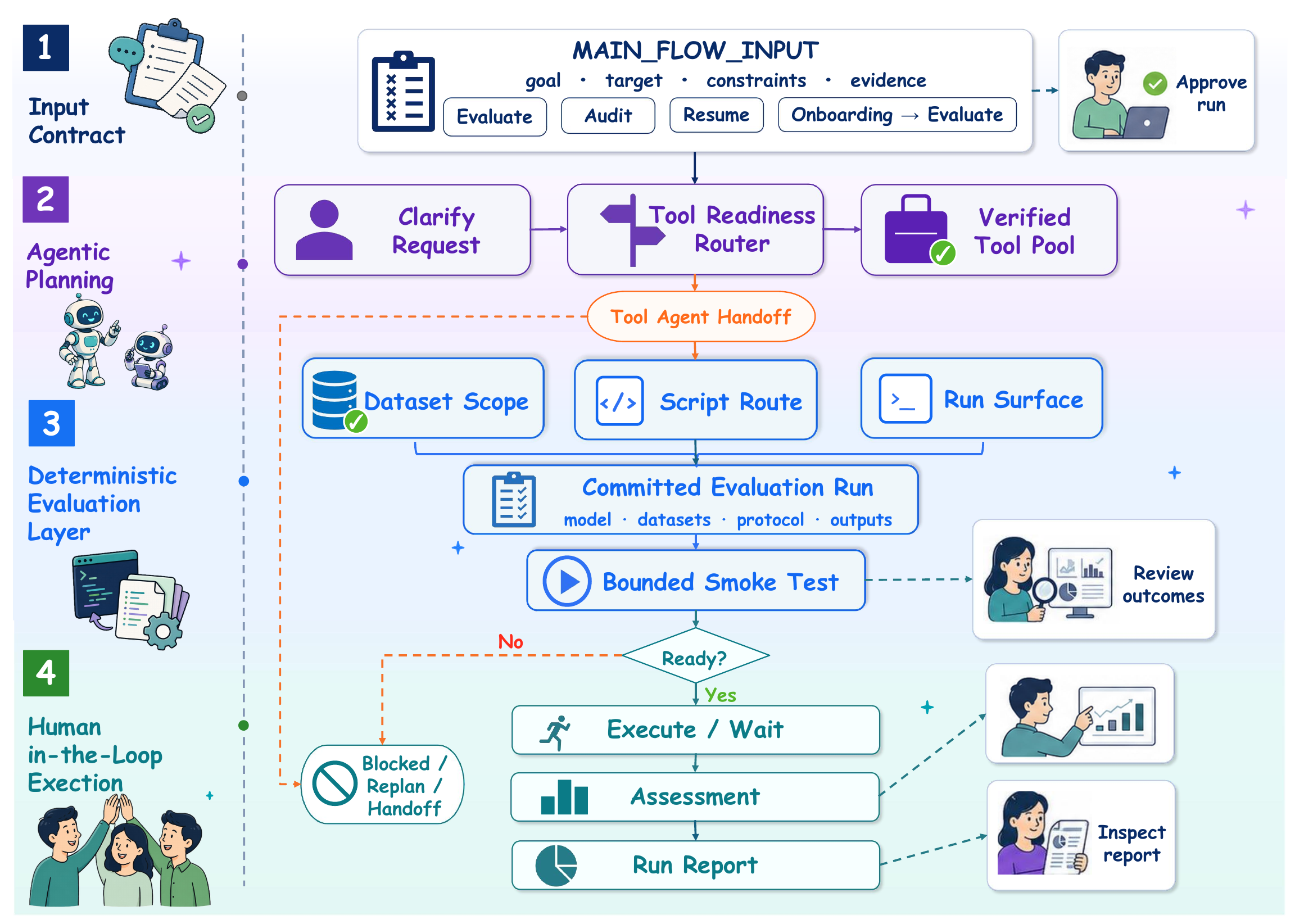}
\caption{Main Agent Workflow from readiness-aware intent resolution to committed execution and auditable run evidence.}
\label{fig:main-agent-workflow}
\end{figure}

The \mainworkflow is the global orchestration layer. Its central output is not a conversational plan but a \emph{committed evaluation run}: a structured record fixing the tool, datasets, inference profile, scoring route, executable entry point, output paths, and expected artifacts before score-bearing execution begins. Figure~\ref{fig:main-agent-workflow} shows the transition from a structured request to readiness routing, protocol commitment, bounded smoke testing, execution, assessment, and reporting.

The structured request records the evaluation goal, target model, allowed tasks and datasets, execution constraints, available evidence, and artifact locations. A declared directory or endpoint is not sufficient evidence of readiness. If the callable basis is absent, broken, or unverified, evaluation is paused and the target is handed to the Tool Agent. The Main Agent does not repair dependencies or redesign wrappers inside an evaluation run.

After readiness is established, the Main Agent resolves three coupled objects: the dataset scope, the script route, and the run surface. The dataset scope records both selected and excluded datasets with capability- or constraint-based reasons. The script route resolves registered preparation, prediction, validation, canonicalization, metric, and reporting nodes. The run surface binds these choices to concrete input/output locations and an executable entry point. Importantly, the result of planning is materialized before the costly run begins: once committed, a model-specific error cannot trigger an unrecorded change of dataset, decoding option, normalization rule, or evaluator. Any such change requires a new contract and a new run identity.

\subsubsection{Inference protocol.}
The inference protocol fixes both hard and soft execution conditions. Hard conditions include accelerator class and topology, runtime revision, precision, and quantization. Soft conditions include task/language modes, temperature, beam or search strength, fallback and retry behavior, prompt/context propagation, and streaming/chunking settings. Parameters are normalized by semantics rather than by name. A field unsupported by a model is recorded as \code{not\_applicable}; it is not silently replaced by a different behavior. Fair comparison therefore means that semantically applicable dimensions are fixed and every exception remains visible.

\subsubsection{Scoring protocol.}
The scoring protocol fixes the complete route
\begin{equation}
\begin{aligned}
\text{Raw Output}
&\rightarrow \text{Schema Validation}
\rightarrow \text{Canonicalization}\\
&\rightarrow \text{Metric Node(s)}
\rightarrow \text{Aggregation}
\rightarrow \text{Report}.
\end{aligned}
\label{eq:scoring-route}
\end{equation}
Each stage is a registered and versioned procedure. The Main Agent may select a route compatible with the task and language, but may not author a new normalization rule or metric during execution. Before a full run, a bounded smoke test verifies the same end-to-end surface on a small workload; a full benchmark is never used as the first integration test.

The smoke test validates the committed surface rather than a simplified surrogate. It checks that the selected dataset record can be read, the verified tool can generate a schema-valid prediction, the canonicalizer accepts that prediction, and the resolved metric node can emit the expected report fields. A failed smoke test produces an explicit blocked state and evidence for replanning or Tool-Agent handoff. It never authorizes the Main Agent to bypass a failing node or to report a partial score under the original contract.

\begin{figure}[!htbp]
\centering
\includegraphics[width=0.92\linewidth]{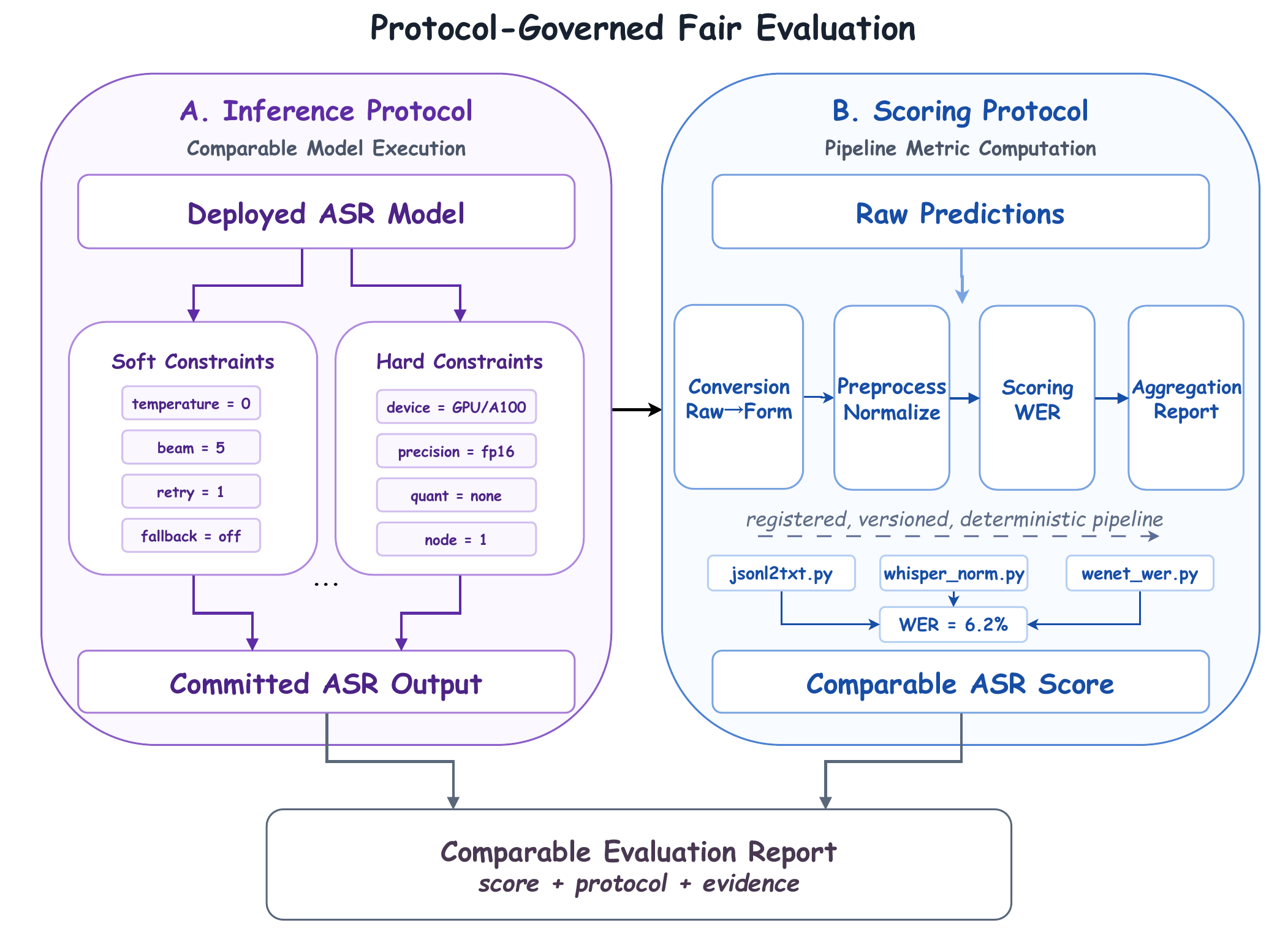}
\caption{Protocol-governed fair evaluation in SURE-EVAL. Comparable results require committed inference and scoring contracts before score-bearing execution.}
\label{fig:protocol-governed-fair-evaluation}
\end{figure}

\subsection{Versioned Speech Evaluation Registry}
\label{sec:registry}

SURE-EVAL currently registers evaluation routes for recognition, translation, speaker-aware processing, generation, conversion, classification, and keyword spotting. Table~\ref{tab:evaluation-registry} provides a compact snapshot; anonymized node manifests and representative run artifacts are included in the supplementary material. A route is represented as a directed sequence of nodes rather than a metric name alone. Every node declares its identifier, implementation version, environment, expected inputs, outputs, and compatible task/language conditions. The run report stores the resolved node versions together with predictions and aggregate metrics, so a later update to a normalizer, ASR frontend, speaker encoder, or MOS predictor cannot silently redefine an earlier result.

\begin{table}[!htbp]
\setlength{\belowcaptionskip}{3pt}
\centering
\scriptsize
\caption{Representative task and metric routes currently integrated in SURE-EVAL. Each arrow denotes a separately versioned node with an explicit input/output contract.}
\label{tab:evaluation-registry}
\begin{tabularx}{\linewidth}{@{}P{27mm}P{45mm}Y@{}}
\toprule
\textbf{Task family} & \textbf{Representative metrics} & \textbf{Versioned route} \\
\midrule
ASR & WER, CER, MER & output extraction $\rightarrow$ language-aware normalization $\rightarrow$ tokenization/alignment $\rightarrow$ aggregation \\
Speech translation & BLEU~\cite{papineni2002bleu}, chrF2~\cite{popovic2015chrf}, XCOMET-XL~\cite{guerreiro2023xcomet}, BLEURT-20~\cite{sellam2020bleurt} & text canonicalization $\rightarrow$ metric-specific model/script $\rightarrow$ aggregation \\
SD / SA-ASR & DER; cpWER and DER & RTTM/transcript validation $\rightarrow$ canonicalization $\rightarrow$ MeetEval scoring~\cite{vonneumann2023meeteval} \\
TTS / VC & CER/WER; WavLM~\cite{chen2022wavlm}, ECAPA-TDNN~\cite{desplanques2020ecapa}, ERes2Net~\cite{chen2023eres2net}; DNSMOS~\cite{reddy2021dnsmos}, WV-MOS~\cite{cooper2022wvmos}, UTMOS~\cite{saeki2022utmos} & audio frontend $\rightarrow$ ASR/SV/MOS nodes $\rightarrow$ task-level aggregation \\
SER / GR / SLU & Accuracy and task-specific F1 & label/output canonicalization $\rightarrow$ classification or answer matching \\
KWS & Accuracy, precision, recall, F1, FRR, FAR & event canonicalization $\rightarrow$ detection matching $\rightarrow$ metric aggregation \\
\bottomrule
\end{tabularx}
\end{table}

Node resolution occurs when the run is committed. The resolved route records the node identifier and revision rather than a mutable ``latest'' alias, and each node validates both the input schema received from its predecessor and the output schema promised to its successor. This is particularly important for composite speech metrics. A TTS content-error score, for example, depends on an audio frontend, an ASR checkpoint, decoding settings, transcript extraction, language-specific normalization, and WER/CER aggregation. Speaker similarity and non-intrusive quality scores likewise depend on the exact encoder or predictor checkpoint. Treating each of these elements as a node exposes where two nominally identical metrics cease to be equivalent.

The registry does not claim that one normalization or evaluator is universally correct. It requires the selected equivalence policy to be explicit, consistently applied to every compared system, and recoverable from the run artifacts. This distinction is essential: standardizing the pipeline means standardizing the declared route and versions, not hiding task-specific choices behind a generic metric label.

\section{Framework Evaluation and Analysis}
\label{sec:experiments}

\subsection{Experimental Setup}
\label{sec:setup}

We evaluate two capabilities of the framework: heterogeneous model onboarding and protocol-governed model evaluation. For onboarding, both conditions use the same Codex agent, configured with GPT-5.5 and the \code{xhigh} reasoning effort~\cite{openaicodex2026,openaigpt552026}, the same initial model specification, the same upstream evidence, and the same execution budget. The \emph{Codex-only} baseline receives the generic goal of constructing and validating a callable model interface. The \emph{Codex + Tool Agent} condition additionally uses SURE-EVAL's context routing, layered memory, task/environment guidance, structured onboarding contract, and four validation gates. A model is counted as \emph{one-shot} only when import, checkpoint loading, bounded inference, and output-contract validation all pass within the first agent invocation, without an additional corrective instruction from the user.

The onboarding set contains 18 public releases: eight dedicated ASR systems, one multi-task audio model, six TTS systems, one VC system, one speaker-diarization system, and one speaker-attributed ASR system. We use the release names and checkpoints listed in Table~\ref{tab:tool-agent-onboarding-coverage}; each row is linked to the corresponding official model card, repository, or paper. The comparison is intended to isolate the effect of workflow scaffolding while holding the underlying coding agent and available evidence fixed.

For the Main Agent experiments, all local inference is executed on a single NVIDIA RTX~4090 with batch size one and no quantization. Checkpoint and runtime revisions are fixed in the run manifests. For ASR decoders exposing the corresponding controls, temperature is fixed to 0, beam size to 1, and fallback/retry behavior is disabled; unsupported controls are recorded as \code{not\_applicable}. Dataset manifests, audio preprocessing, segmentation, output extraction, normalization, and metric-node versions are identical across compared models. The ASR evaluation covers LibriSpeech test-clean/test-other~\cite{panayotov2015librispeech}, AISHELL-1~\cite{bu2017aishell1}, WenetSpeech meeting/net~\cite{zhang2022wenetspeech}, GigaSpeech~\cite{chen2021gigaspeech}, and SlideSpeech~\cite{wang2023slidespeech}. TTS systems use the same prompts, targets, and subset definitions from Seed-TTS-Eval~\cite{anastassiou2024seedtts}, covering zh and en; model-specific generation controls are frozen before execution and recorded in their inference contracts.

\subsection{Tool Agent Onboarding Coverage}
\label{sec:onboarding-results}

Table~\ref{tab:tool-agent-onboarding-coverage} compares first-pass completion under the Codex-only baseline and the same Codex agent equipped with the SURE-EVAL Tool Agent Workflow. The full workflow completes all 18 releases in one shot. The baseline completes 12, but does not pass all four gates in the first invocation for Parakeet RNNT 1.1B, Whisper large-v3-turbo, Fun-CosyVoice3-0.5B-2512, dots.tts, DiariZen, or Sortformer.

\begin{table}[!htbp]
\setlength{\belowcaptionskip}{3pt}
\centering
\scriptsize
\caption{One-shot onboarding with a Codex-only baseline versus the same Codex agent equipped with the SURE-EVAL Tool Agent Workflow. A pass requires all four validation gates to succeed in the first invocation.}
\label{tab:tool-agent-onboarding-coverage}
\setlength{\tabcolsep}{3.2pt}
\renewcommand{\arraystretch}{1.04}
\begin{tabularx}{\linewidth}{@{}P{58mm}P{15mm}>{\centering\arraybackslash}P{23mm}>{\centering\arraybackslash}Y@{}}
\toprule
\textbf{Release / checkpoint} & \textbf{Task} & \textbf{Codex only} & \textbf{Codex + Tool Agent} \\
\midrule
Parakeet RNNT 1.1B (en)~\cite{parakeetmodel} & ASR & \fail & \pass \\
SenseVoiceSmall~\cite{sensevoicemodel} & ASR & \pass & \pass \\
Qwen3-ASR-1.7B~\cite{qwen3asrmodel} & ASR & \pass & \pass \\
Whisper large-v3-turbo~\cite{whisperturbomodel} & ASR & \fail & \pass \\
MOSS-Transcribe-preview-2B~\cite{mosstranscribemodel} & ASR & \pass & \pass \\
Granite-Speech-4.1-2B~\cite{granitespeechmodel} & ASR & \pass & \pass \\
X-ASR-zh-en~\cite{xasrmodel} & ASR & \pass & \pass \\
FireRedASR-LLM-L~\cite{fireredmodel} & ASR & \pass & \pass \\
Kimi-Audio-7B-Instruct~\cite{kimiaudiomodel} & Multi & \pass & \pass \\
F5-TTS~\cite{chen2025f5tts} & TTS & \pass & \pass \\
IndexTTS2~\cite{zhou2025indextts2} & TTS & \pass & \pass \\
Fun-CosyVoice3-0.5B-2512~\cite{cosyvoice3model} & TTS & \fail & \pass \\
Qwen3-TTS-12Hz-1.7B-Base~\cite{qwen3ttsmodel} & TTS & \pass & \pass \\
VoxCPM2~\cite{voxcpm2model} & TTS & \pass & \pass \\
dots.tts~\cite{lian2026dotstts} & TTS & \fail & \pass \\
X-VC~\cite{zheng2026xvc} & VC & \pass & \pass \\
DiariZen~\cite{han2025diarizen} & SD & \fail & \pass \\
Sortformer~\cite{park2025sortformer} & SA-ASR & \fail & \pass \\
\midrule
\textbf{One-shot total} & & \textbf{12/18} & \textbf{18/18} \\
\bottomrule
\end{tabularx}
\end{table}

The six baseline failures are distributed across recognition, synthesis, diarization, and speaker-attributed recognition rather than concentrated in one task family. Their releases expose different sources of onboarding ambiguity, including backend-specific dependency stacks, nonstandard checkpoint layouts, model-specific generation APIs, and outputs that require task-aware schema mapping. The result therefore does not merely reflect a single repository template learned by the agent. With the same underlying Codex model and evidence held fixed, the structured workflow converts these heterogeneous cases into the same verified handoff interface on the first invocation.

The comparison also clarifies what is and is not claimed. Table~\ref{tab:tool-agent-onboarding-coverage} measures first-pass completion of a bounded onboarding task, not downstream model quality and not repeated-run statistical reliability. Its methodological value is to show that selective context, role-aligned memory, explicit unresolved fields, and ordered validation gates prevent common integration failures from escaping into the benchmark stage. In particular, a package import or nonempty model response is insufficient: every full-workflow pass also satisfies the output contract required by deterministic evaluation.

\subsection{Unified ASR Evaluation}
\label{sec:asr-results}

Table~\ref{tab:main-agent-asr-results} reports the ASR matrix produced after the Main Agent commits the common hardware, inference, dataset, output, and scoring contracts. The table is intentionally not interpreted as an unconstrained leaderboard: some checkpoints are English-only or have limited Chinese support, and the protocol does not silently select a more favorable language mode. For English-only checkpoints, Mandarin conditions are reported as unsupported and marked with ``-'' rather than retaining non-meaningful CER values above 100\%. In contrast, multilingual systems such as Qwen3-ASR and SenseVoice remain usable across both English and Mandarin datasets.

\begin{table}[!htbp]
\setlength{\belowcaptionskip}{3pt}
\centering
\scriptsize
\caption{ASR evaluation under the committed SURE-EVAL protocol. Values are WER for English and CER for Mandarin, in percent (lower is better). English-only systems do not support Mandarin conditions; their corresponding CER cells are filled with ``-'' rather than reporting non-meaningful values.}
\label{tab:main-agent-asr-results}
\setlength{\tabcolsep}{2pt}
\begin{tabularx}{\linewidth}{@{}P{36mm}*{7}{>{\centering\arraybackslash}X}@{}}
\toprule
\textbf{Model} &
\textbf{LS clean} &
\textbf{LS other} &
\textbf{AI-1} &
\textbf{Wenet meet.} &
\textbf{Wenet net} &
\textbf{Giga} &
\textbf{Slide} \\
\midrule
\code{Parakeet-RNNT-1.1B}~\cite{parakeetmodel} & 1.45 & \textbf{2.48} & - & - & - & 9.92 & 8.54 \\
\code{SenseVoiceSmall}~\cite{sensevoicemodel} & 3.13 & 7.19 & 3.00 & 7.39 & 7.31 & 12.79 & 8.97 \\
\code{Qwen3-ASR-1.7B}~\cite{qwen3asrmodel} & 1.63 & 3.41 & \textbf{1.58} & \textbf{5.78} & \textbf{4.98} & 8.75 & \textbf{7.09} \\
\code{Whisper-v3-turbo}~\cite{whisperturbomodel} & 1.95 & 3.97 & 7.23 & 21.94 & 14.03 & 9.98 & 8.76 \\
\code{MOSS-Transcribe-2B}~\cite{mosstranscribemodel} & 1.51 & 3.89 & 79.91 & 84.04 & 43.41 & \textbf{8.24} & 7.68 \\
\code{Granite-Speech-4.1-2B}~\cite{granitespeechmodel} & \textbf{1.32} & 2.68 & - & - & - & 10.29 & 8.73 \\
\bottomrule
\end{tabularx}
\vspace{2pt}
\begin{minipage}{0.98\linewidth}
\footnotesize
LS denotes LibriSpeech~\cite{panayotov2015librispeech} and AI-1 denotes AISHELL-1~\cite{bu2017aishell1}. Wenet, Giga, and Slide denote WenetSpeech~\cite{zhang2022wenetspeech}, GigaSpeech~\cite{chen2021gigaspeech}, and SlideSpeech~\cite{wang2023slidespeech}. Bold values mark the best supported score in each column. All cells use the same registered manifests and language-specific normalization/scoring nodes.
\end{minipage}
\end{table}

The matrix illustrates why deployment and scoring must be separated from capability claims. On LibriSpeech, several specialized English systems achieve low WER, whereas their Chinese outputs are not meaningful. Qwen3-ASR achieves the most consistent cross-lingual coverage among the evaluated rows, while SenseVoice remains competitive on Mandarin despite weaker English results. These conclusions are traceable to the same scoring route; they are not confounded by per-model normalizers or selectively chosen evaluators. The run records also preserve unsupported-language behavior, which is operationally relevant for model selection but is often omitted from paper-specific result tables.

\subsection{Unified TTS Evaluation}
\label{sec:tts-results}

TTS requires a composite evaluation rather than one scalar. Table~\ref{tab:main-agent-tts-results} therefore reports content error, three speaker-similarity estimators, and three non-intrusive quality predictors. Every generated waveform is processed by the same registered audio frontends and fixed model versions. This design avoids the common situation in which two TTS systems are compared using different ASR recognizers, speaker encoders, or MOS predictors.

\begin{table}[!htbp]
\setlength{\belowcaptionskip}{3pt}
\centering
\footnotesize
\caption{TTS evaluation under the committed SURE-EVAL protocol on the zh/en subsets of Seed-TTS-Eval~\cite{anastassiou2024seedtts}.}
\label{tab:main-agent-tts-results}
\setlength{\tabcolsep}{0.9pt}
\begin{tabular}{@{}P{23mm}P{12mm}*{7}{c}@{}}
\toprule
\textbf{Model} & \textbf{Split} & \mbox{\textbf{C-Err.}} & \mbox{\textbf{S-ECA1}} & \mbox{\textbf{S-ECA2}} & \mbox{\textbf{S-ER2}} & \mbox{\textbf{M-DNS}} & \mbox{\textbf{M-WV}} & \mbox{\textbf{M-UT}} \\
\midrule
\multirow{2}{=}{\code{F5-TTS}~\cite{chen2025f5tts}} & \code{zh} & 1.54 & 0.78 & 0.80 & 0.84 & 3.35 & 3.38 & 2.95 \\
 & \code{en} & 1.31 & 0.67 & \textbf{0.75} & 0.81 & 3.22 & 4.07 & 3.67 \\
\addlinespace[2pt]
\multirow{2}{=}{\code{IndexTTS2}~\cite{zhou2025indextts2}} & \code{zh} & 1.09 & 0.78 & 0.79 & 0.86 & 3.30 & 3.60 & 3.00 \\
 & \code{en} & 1.18 & 0.70 & \textbf{0.75} & \textbf{0.86} & 3.07 & 3.99 & 3.65 \\
\addlinespace[2pt]
\multirow{2}{=}{\code{CosyVoice3}~\cite{cosyvoice3model}} & \code{zh} & 1.19 & 0.79 & \textbf{0.81} & \textbf{0.87} & 3.34 & 3.54 & 3.32 \\
 & \code{en} & 1.53 & 0.69 & 0.74 & 0.80 & 3.20 & 4.07 & 3.96 \\
\addlinespace[2pt]
\multirow{2}{=}{\code{Qwen3-TTS}~\cite{qwen3ttsmodel} (\code{xvec=T})} & \code{zh} & \textbf{0.87} & 0.74 & 0.74 & 0.76 & \textbf{3.38} & \textbf{3.64} & \textbf{3.53} \\
 & \code{en} & \textbf{0.92} & 0.61 & 0.65 & 0.74 & \textbf{3.28} & 4.32 & \textbf{4.25} \\
\addlinespace[2pt]
\multirow{2}{=}{\code{Qwen3-TTS}~\cite{qwen3ttsmodel} (\code{xvec=F})} & \code{zh} & 1.03 & 0.77 & 0.75 & 0.80 & 3.37 & \textbf{3.64} & 3.51 \\
 & \code{en} & 1.11 & 0.71 & 0.70 & 0.77 & 3.22 & \textbf{4.33} & 4.17 \\
\addlinespace[2pt]
\multirow{2}{=}{\code{VoxCPM2}~\cite{voxcpm2model}} & \code{zh} & 0.98 & 0.78 & 0.74 & 0.80 & 3.26 & 3.43 & 2.98 \\
 & \code{en} & 1.24 & 0.72 & 0.70 & 0.76 & 3.09 & 4.04 & 3.80 \\
\addlinespace[2pt]
\multirow{2}{=}{\code{dots.tts}~\cite{lian2026dotstts}} & \code{zh} & 1.04 & \textbf{0.80} & 0.79 & 0.82 & 3.30 & 3.53 & 3.10 \\
 & \code{en} & 1.11 & \textbf{0.76} & \textbf{0.75} & 0.78 & 3.18 & 4.03 & 3.81 \\
\bottomrule
\end{tabular}
\vspace{2pt}
\begin{minipage}{0.98\linewidth}
\footnotesize
C-Err. is CER for \code{zh} and WER for \code{en} ($\downarrow$). Bold values mark the best score within each split and metric column. S-ECA1 loads the Seed-TTS-Eval \href{https://huggingface.co/lmzjms/wavlm-large/blob/f1c40859a86883dac660f314165a2ea45a8e4dab/wavlm_large_finetune.pth}{\code{wavlm\_large\_finetune.pth}} checkpoint~\cite{anastassiou2024seedtts}; S-ECA2 uses \href{https://huggingface.co/speechbrain/spkrec-ecapa-voxceleb}{\code{speechbrain/spkrec-ecapa-voxceleb}}~\cite{desplanques2020ecapa}; and S-ER2 uses ERes2Net~\cite{chen2023eres2net} ($\uparrow$). M-DNS, M-WV, and M-UT use DNSMOS~\cite{reddy2021dnsmos}, WV-MOS~\cite{cooper2022wvmos}, and UTMOS~\cite{saeki2022utmos} ($\uparrow$). For Qwen3-TTS, \code{xvec=T/F} denotes \code{x\_vector\_mode=true/false}.
\end{minipage}
\end{table}

The unified view exposes trade-offs that a single paper-specific metric can hide. Qwen3-TTS with the default \code{x\_vector\_mode=true} obtains the lowest content error on the two standard zh/en subsets and the strongest UTMOS scores, while the \code{x\_vector\_mode=false} run improves all three speaker-similarity estimates and yields the highest English WV-MOS. IndexTTS2 and CosyVoice3 remain strong on ERes2Net similarity, and dots.tts is also competitive: it achieves the strongest S-ECA1 scores on both splits, ties for the best English S-ECA2 score, and keeps content error close to the leading systems. The three speaker encoders and three quality predictors do not induce identical rankings, confirming that the identity, version, and mode of each evaluation node are part of the scientific claim rather than interchangeable implementation details.

\subsection{Protocol Analysis: Reported Results versus the Unified Contract}
\label{sec:protocol-analysis}

We use Table~\ref{tab:reported-vs-unified} as a compact system-level protocol ablation. For three TTS systems evaluated on the same nominal Seed-TTS zh/en subsets, we replace the paper-specific execution and scoring setup with the committed SURE-EVAL contract. This is deliberately a multi-factor ablation: model checkpoint selection, inference parameters, prompt/reference handling, stochastic generation, ASR frontend, normalization, and aggregation may differ when the original paper does not expose an executable node-level contract.

\begin{table}[!htbp]
\setlength{\belowcaptionskip}{3pt}
\raggedright
\scriptsize
\caption{Paper-reported versus unified content error on Seed-TTS zh/en (\%). $\Delta$ is unified minus reported.}
\label{tab:reported-vs-unified}
\setlength{\tabcolsep}{3pt}
\noindent\begin{tabular*}{\textwidth}{@{}P{37mm}@{\extracolsep{\fill}}ccc ccc@{}}
\toprule
& \multicolumn{3}{c}{\textbf{Seed-TTS zh}} & \multicolumn{3}{c}{\textbf{Seed-TTS en}} \\
\cmidrule(lr){2-4}\cmidrule(lr){5-7}
\textbf{Model / reported configuration} & \textbf{Reported} & \textbf{Unified} & $\bm{\Delta}$ & \textbf{Reported} & \textbf{Unified} & $\bm{\Delta}$ \\
\midrule
F5-TTS (32 NFE)~\cite{chen2025f5tts} & 1.56 & 1.54 & -0.02 & 1.83 & 1.31 & -0.52 \\
Qwen3-TTS-12Hz-1.7B-Base~\cite{hu2026qwen3tts} & 0.77 & 0.87 & +0.10 & 1.24 & 0.92 & -0.32 \\
dots.tts (base)~\cite{lian2026dotstts} & 0.96 & 1.04 & +0.08 & 1.34 & 1.11 & -0.23 \\
\bottomrule
\end{tabular*}
\end{table}

Table~\ref{tab:reported-vs-unified} shows that the reported-versus-unified gap is model- and language-dependent rather than uniformly stricter or more permissive. On zh, the unified scores are close to the reported values, with shifts from -0.02 to +0.10 points; on en, all three unified scores are lower by 0.23--0.52 points. These differences reflect interactions among the effective model, generation protocol, reference construction, ASR frontend, normalization, and aggregation route. We do not interpret the original results as incorrect; they describe different, sometimes incompletely specified evaluation claims. The value of SURE-EVAL is that the corresponding conditions are made explicit, versioned, and rerunnable, so the observed gap can be audited rather than attributed vaguely to ``reproduction error.''

\section{Conclusion and Outlook}
\label{sec:conclusion}

We presented SURE-EVAL, a systematic and unified agentic framework that binds a verified callable model, inference protocol, dataset, output contract, versioned scoring route, and run evidence into one executable evaluation claim. Across 18 releases, the Codex-only baseline completes 12 one-shot onboardings, whereas the same Codex agent with the Tool Agent Workflow completes all 18. Unified ASR/TTS evaluation and the reported-versus-unified analysis further show that hidden execution and scoring choices can materially change the interpretation of a result. Future work will broaden the model, task, dataset, and metric-node registry and support community services for paper-result reproduction and protocol-aware model comparison, with the goal of making speech research more transparent, comparable, and reproducible.

\begin{small}
\par\medskip\noindent\textbf{Acknowledgments.}
This work has been supported by the China NSFC Project (No. 92370206), the Key Research and Development Program of Jiangsu Province, China (Grant No. BF2025029) and the Joint Fund Project of Shanghai Commercial Aircraft Systems Engineering Science.
\par\smallskip\noindent\textbf{Disclosure of Interests.}
Yihua Zhou, Qiang Zhou, and Feng Lu are affiliated with AISpeech Co., Ltd. Junhao Du, Yixuan Wang, Bowen Wang, Ruichen Sun, and Chenghao Wang contributed to this work during internships at AISpeech Co., Ltd. The authors have no other competing interests to declare that are relevant to the content of this article.
\end{small}


\clearpage
\bibliographystyle{splncs04}
\bibliography{references}

\end{document}